\documentclass[traditabstract]{aa}

\usepackage{epsfig}
\usepackage{graphicx}
\usepackage{latexsym}
\usepackage{amssymb}
\usepackage{amsmath}
\usepackage{longtable}
\usepackage{color}
\usepackage{txfonts}
\usepackage{appendix}
\usepackage{ulem}
\usepackage{footnote}
\usepackage{caption}
\usepackage{subcaption}

\bibstyle{aa}

\newcommand{\mum}{\mathrm{\mu m}}

\newcommand{\surfb}{$\mathrm{W}\ \mathrm{m^{-2}}\ \mathrm{sr^{-1}}\ \mathrm{Hz^{-1}}$}
\newcommand{\vghz}{$\mathrm{V}\,\mathrm{GHz^{-1}}$}
\newcommand{\rtel}{$R_{tel}$}

\usepackage{natbib,twoopt}
\usepackage[breaklinks=true]{hyperref} %% to avoid \citeads line fills
\bibpunct{(}{)}{;}{a}{}{,} %% natbib format for A&A and ApJ
\makeatletter
\newcommandtwoopt{\citeads}[3][][]{\href{http://adsabs.harvard.edu/abs/#3}%
{\def\hyper@linkstart##1##2{}%
\let\hyper@linkend\@empty\citealp[#1][#2]{#3}}}
\newcommandtwoopt{\citepads}[3][][]{\href{http://adsabs.harvard.edu/abs/#3}%
{\def\hyper@linkstart##1##2{}%
\let\hyper@linkend\@empty\citep[#1][#2]{#3}}}
\newcommandtwoopt{\citetads}[3][][]{\href{http://adsabs.harvard.edu/abs/#3}%
{\def\hyper@linkstart##1##2{}%
\let\hyper@linkend\@empty\citet[#1][#2]{#3}}}
\newcommandtwoopt{\citeyearads}[3][][]%
{\href{http://adsabs.harvard.edu/abs/#3}
{\def\hyper@linkstart##1##2{}%
\let\hyper@linkend\@empty\citeyear[#1][#2]{#3}}}
\makeatother
\begin{document}
\sloppy

\title{Observing Extended Sources with the \textsl{Herschel} SPIRE Fourier Transform Spectrometer}

\author{R. Wu\inst{1},
E.~T.~Polehampton\inst{2,3}, 
M. Etxaluze\inst{4}, 
G. Makiwa\inst{3}, 
D.~A.~Naylor\inst{3},
C. Salji\inst{2,5,6}, 
B.~M.~Swinyard\inst{2,6}, 
M. Ferlet\inst{2}, 
M.~H.~D. van der Wiel\inst{3},
A.~J.~Smith\inst{7},  
T. Fulton\inst{3,7},
M.~J.~Griffin\inst{8},
J.-P. Baluteau\inst{9},
D. Benielli\inst{9},
J. Glenn\inst{10},
R. Hopwood\inst{11},
P. Imhof\inst{3,7},
T. Lim\inst{2}, 
N. Lu\inst{12},
P. Panuzzo\inst{1},
C. Pearson\inst{2,13}, 
S. Sidher\inst{2}, 
I. Valtchanov\inst{14}}

\institute{
%1
Laboratoire AIM, CEA/DSM - CNRS - Irfu/Service d'Astrophysique, CEA Saclay, 91191 Gif-sur-Yvette, France
\and
%2
RAL Space, Rutherford Appleton Laboratory, Didcot OX11 0QX, UK
\and
%3
Institute for Space Imaging Science, Department of Physics \& Astronomy, University of Lethbridge, Lethbridge, AB T1K3M4, Canada
\and
%4
Centro de Astrobiolog\'ia (CSIC/INTA), Ctra. de Torrej\'on a Ajalvir, km 4, 28850 Torrej\'on de Ardoz, Madrid, Spain
\and
%5
Cambridge University, Cavendish Laboratory and the Kavli Institute for Cosmology, Cambridge CB3 0HA, United Kingdom
\and
%6
University College London, Deptartment of Physics and Astronomy, London WC1E 6BT, United Kingdom
\and
%7
Blue Sky Spectroscopy, 9 / 740 4 Ave S, Lethbridge, Alberta, T1J 0N9, Canada
\and
%8
Cardiff University, The Parade, Cardiff, UK
\and
%9
Laboratoire d'Astrophysique de Marseille - LAM, Universit\'e d'Aix-Marseille \& CNRS, UMR7326, 38 rue F. Joliot-Curie, 13388 Marseille Cedex 13, France 
\and
%10
Center for Astrophysics and Space Astronomy, 389-UCB, University of Colorado, Boulder, CO 80303, USA
\and
%11
Physics Department, Imperial College London, South Kensington Campus, SW7 2AZ, UK
\and
%12
NASA Herschel Science Centre, IPAC, Pasadena, California, USA
\and
%13
Department of Physical Sciences, The Open University, Milton Keynes MK7 6AA, UK
\and
%14
European Space Astronomy Centre, Herschel Science Centre, ESA, 28691 Villanueva de la Ca\~nada, Spain
}
\authorrunning{Ronin Wu et al.}
\abstract{The Spectral and Photometric Imaging Receiver (SPIRE) on the European Space Agency's \textsl{Herschel} Space Observatory utilizes a pioneering design for its imaging spectrometer in the form of a Fourier Transform Spectrometer (FTS). The standard FTS data reduction and calibration schemes are aimed at objects with either a spatial extent much larger than the beam size or a source that can be approximated as a point source within the beam. However, when sources are of intermediate spatial extent, neither of these calibrations schemes is appropriate and both the spatial response of the instrument and the source's light profile must be taken into account and the coupling between them explicitly derived. To that end, we derive the necessary corrections using an observed spectrum of a fully extended source with the beam profile and the source's light profile taken into account. We apply the derived correction to several observations of planets and compare the corrected spectra with their spectral models to study the beam coupling efficiency of the instrument in the case of partially extended sources. We find that we can apply these correction factors for sources with angular sizes up to $\theta_{D}\sim17''$. We demonstrate how the angular size of an extended source can be estimated using the difference between the sub-spectra observed at the overlap bandwidth of the two frequency channels in the spectrometer,  at $959<\nu<989\,\mathrm{GHz}$. Using this technique on an observation of Saturn, we estimate a size of 17.2'', which is 3\% larger than its true size on the day of observation. Finally, we show the results of the correction applied on observations of a nearby galaxy, M82, and the compact core of a Galactic molecular cloud, Sgr B2.}

\keywords{
Instrumentation:spectrographs
-
Methods: analytical
-
Methods: data analysis
-
Techniques:spectroscopic
}

\date{Received; accepted}
\maketitle

%%%%%%%%%%%%%%%%%%%%%%%%%%%%%%%%%%%%%%%%%%%%
%   NOTE
%%%%%%%%%%%%%%%%%%%%%%%%%%%%%%%%%%%%%%%%%%%%

% Section 1.
\section{Introduction}\label{intro}

	%Section 1, Paragraph 1
	\indent\par{
        Accurate calibration of astronomical observations is normally achieved by using standard sources whose flux density is well known from other telescopes and/or is well modelled. However, this limits the applicability of the calibration to those sources which have the same spatial extent within the telescope beam as the standards. A correction using simple filling factors is often used for real astronomical objects that have a non-negligible size with respect to the telescope beam. This paper addresses the problem of sources that are partially extended with respect to the beam of the \textsl{Herschel} SPIRE instrument.
	}

	%Section 1, Paragraph 2
	\par{
	The Spectral and Photometric Imaging Receiver (SPIRE) is one of three focal plane instruments on board the ESA \textsl{Herschel} Space Observatory~\citep{pilbratt10}. It contains an imaging photometric camera and an imaging Fourier Transform Spectrometer (FTS).  Both sub-instruments use arrays of bolometric detectors operating at $\sim$300~mK \citep{turner2001} with feedhorn focal plane optics giving sparse spatial sampling over an extended field of view \citep{dohlen2000}. The FTS is composed of two bolometer arrays with overlapping bands. The SPIRE Long Wavelength Array~(SLW) covers $447<\nu<990\,\mathrm{GHz}$, and the SPIRE Short Wavelength Array~(SSW) covers $958<\nu<1546\,\mathrm{GHz}$. The SLW and SSW contain 19 and 37 hexagonally packed detectors separated by 51'' and 33'' respectively. The FTS can observe requested targets in single-pointed or raster mode with sparse, intermediate, or full sampling by moving the SPIRE internal beam steering mirror to 1, 4, or, 16 jiggle positions. The three available sampling mode settings result in observations with spatial points separated by approximately $2$, $1$, and, $1/2$ beams respectively. A more detailed description of the instrument design can be found in~\citet{griffin2010} and more details of the observing modes in the \citet{observersmanual}.
	}
	% Section 1, Paragraph 3
	\par{
	The standard SPIRE FTS pipeline provides data calibrated with two extreme geometrical assumptions: either uniformly extended emission over a region much larger than the beam, or truly point-like emission centered on the optical axis. This calibration scheme was initially presented in~\citet{swinyard2010} and the FTS pipeline described in \citet{fulton2010}. More recent updates are described in the \citet{observersmanual} and \citet{swinyard-aa-2013}. However, neither of these assumptions fits with real astronomical sources, which often have a complex morphology between the two assumed cases.
	}

	% Section 1, Paragraph 4
	\par{
	In this paper, we calculate a correction for extended sources based on a model of their distribution, and knowledge of the beam profile shape for single-pointed sparsely sampled observations. In Section~\ref{sec_beam}, we summarize the current understanding of the FTS beam profile and its dependence on frequency and in Section~\ref{sec_calibration}, we summarize the important points of the FTS pipeline calibration scheme. We then calculate the efficiency factors necessary to correct the standard calibration for a semi-extended source in Section~\ref{sec_semi} and test this with two real observations in Section~\ref{sec_tests}. A discussion of this work is presented in Section~\ref{discussion}. For the convenience of description, in the following context, we refer to a uniformly distributed extended source as an ``extended source'', and an extended source with spatially-dependent distribution as a ``semi-extended source''. 
	}

% Section 2
\section{Beam Profile}\label{sec_beam}

	% Figure 1: Beam size v.s. frequency in different modes.
	\begin{figure}
		\centering
		\includegraphics[width=\hsize]{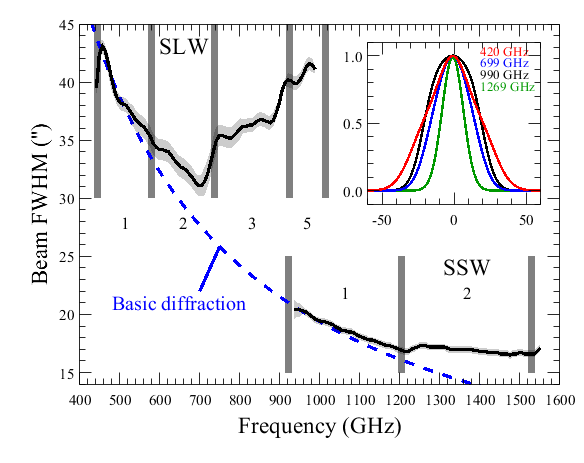}
		\caption{The SPIRE FTS spectral beam size as measured by \citet{makiwa-prep-2013}. The shaded region shows 3$\sigma$ errors. The frequencies at which the waveguide modes are enabled and the number of modes expected to be present (from theory) in each range is also shown, along with the expected size due to single--mode diffraction (dashed blue line). The inset plot shows the shape of the profile at 3 sample frequencies from SLW (420, 699, 990 GHz) and one frequency from SSW (1269 GHz). The horizontal axis of the inset plot is in the unit of arcsec.}
		\label{fig_beam_modes}
	\end{figure}

	% Section 2, Paragraph 1
	\indent\par{The SPIRE FTS detector arrays use feedhorns to couple radiation to the bolometric sensors \citep{turner2001}. Each feedhorn consists of a conical concentrator in front of a circular section waveguide designed to act as a spatial filter, allowing only certain electromagnetic modes to propagate along the waveguide. In most radio and sub-millimeter instruments, the waveguides are sized such that only a single electromagnetic mode is propagated and the resulting spatial response is a well controlled beam that is mathematically well described by a Gaussian profile \citep{martin-ieeetrans-1993}. However, the requirement that the SPIRE FTS cover a large instantaneous bandwidth necessitated the use of multi-moded feedhorns, whose spatial response is determined by the superposition of a finite number of modes that are enabled at specific frequencies. The frequencies at which the higher order modes are enabled are determined by the diameter of the waveguide~\citep{chattopadhyay-ieeetrans-2003, murphy-irphys-1991}.}
	
	% Section 2, Paragraph 2
	\par{One can, in principle, derive the instrument beam pattern by time reversed propagation of the known feedhorn modes through the SPIRE optical system. However, modelling a complex system like the SPIRE FTS, consisting of 18 mirrors, 2 beam splitters, several filters, a dichroic, a lens and an undersized pupil stop, is impractical. Moreover, some clipping of the divergent beam in the two arms of the FTS is inevitable since the location of the intermediate pupil image changes as the spectrometer is scanned~\citep{caldwell2000}. Thus, the efficiency with which each electromagnetic mode couples to the incoming radiation pattern from a source of finite size is not amenable to direct calculation, being dependent on the source distribution. The frequency dependent coupling between a source of a given size and the FTS beam is therefore best determined experimentally using sources of known spatial distribution and flux. This technique has been employed by \citet{makiwa-prep-2013} who used spectral scans of Neptune as the planet was raster scanned across the entrance aperture. The resulting map covered an angular extent equivalent to the third dark Airy ring at 1500 GHz and approximately the first dark Airy ring for frequencies below 510 GHz.}

	% Section 2, Paragraph 3
	\par{When designing sub-millimeter instruments it is common practice to consider the beam as a superposition of a set of discrete orthonormal modes, which are best represented by Hermite-Gaussian (HG) functions~\citep{martin-ieeetrans-1993}. Figure~\ref{fig_beam_modes} shows the frequency dependent beam profile derived by~\citet{makiwa-prep-2013} using HG decomposition, and, for comparison, the FWHM expected from basic diffraction (based on an Airy pattern with effective mirror diameter $3.287~\mathrm{m}$). Also indicated in Figure~\ref{fig_beam_modes} is the number of modes expected to be present given the waveguide diameters. It is clear that, where the waveguide is single-moded (i.e. at frequencies below $\sim580\,\mathrm{GHz}$ in the SLW band and below $\sim1205\,\mathrm{GHz}$ in the SSW band), the instrument is essentially diffraction limited. As further modes propagate in the feedhorn their superposition leads to larger beam sizes relative to the optical diffraction limit. Within the noise limits of the available Neptune data, the SSW band is best represented by a Gaussian function, whereas the SLW band requires the first three radially symmetric HG functions~\citep{makiwa-prep-2013}. These authors also provide the recipe for constructing full radially symmetric beam profiles at any frequency.}

	% Figure 2: Level-1 and Level-2 data for different source sizes
	\begin{figure*}
		\centering
		\includegraphics[width=0.8\textwidth]{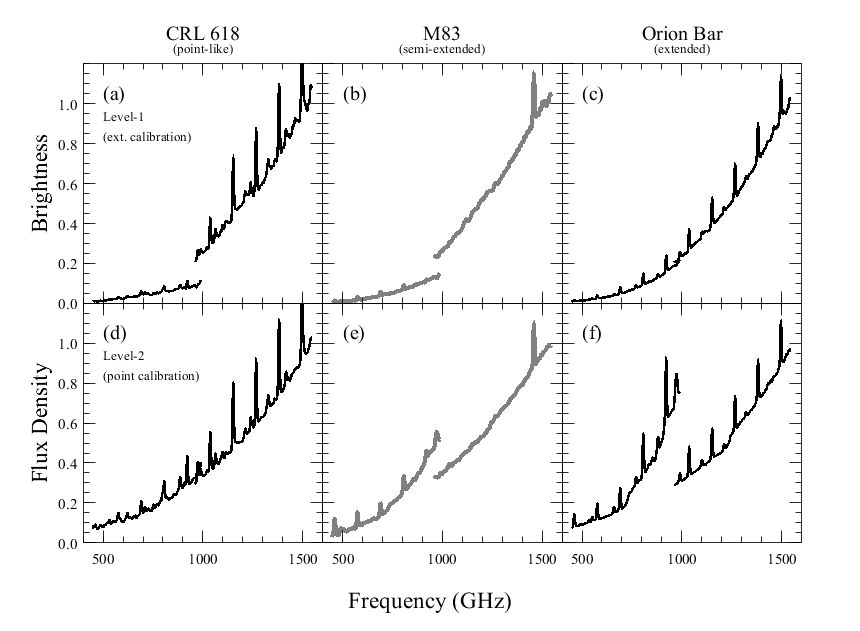}
		\caption{Spectra for a point source (CRL 618, left), semi-extended source (M83, center) and fully extended source (Orion Bar, right) after extended calibration (top row) and point source calibration (bottom row). The spectra have been smoothed. The flux density and brightness are in normalized units.}
		\label{fig_point_ext}
	\end{figure*}

% Section 3
\section{Extended and Point Source Calibration of the FTS}\label{sec_calibration}

	% Section 3, Paragraph 1
	\indent\par{
	In this section, we briefly summarize the standard pipeline calibration scheme for the FTS. The instrument measures an interferogram pattern when observing a given astronomical source, in units of detector voltage. The interferogram is then converted to a spectrum in units of volts per GHz~(\vghz) via a Fourier transform \citep{fulton2010}. The resulting spectrum, in voltage density, is represented by $V_{\mathrm{obs}}$.
	}

	% Section 3, Paragraph 2
	\par{
	The standard pipeline processes all non-mapping observations from voltage density to physical units in two stages. Firstly, the \textsl{Herschel} telescope is used as a calibrator to convert voltage density to {\it surface brightness}, $I_{\mathrm{ext}}$, which is appropriate for sources uniformly extended in the beam, and provides the `Level-1' product. Secondly, a point-source calibration is applied, derived using the planet Uranus, and convert $I_{\mathrm{ext}}$ to {\it flux density}, $F_{\mathrm{point}}$, providing the `Level-2' product. The observer is then able to choose which of Level-1 or Level-2, if either, matches the reality of their source.
	}

	% Section 3.1
	\subsection{Level-1: Extended-Source Calibration}\label{subsec_ex_calibration}

		% Section 3.1, Paragraph 1
		\indent\par{
		As described above, the Level-1 spectrum output by the FTS pipeline is calibrated assuming that the source is uniformly extended within the beam. The calibration is based on observations of a dark region of sky, i.e. where the detectors should only be receiving thermal radiation from the \textsl{Herschel} telescope and SPIRE instrument. Models of the telescope and instrument emission are constructed using onboard temperature sensors and the telescope mirror emissivity as measured on the ground \citep{fischer}, as described in the~\citet{observersmanual}. The spectral response applicable to extended sources is then defined as,

		% Equation 1
		\begin{equation}
			\label{eq1a}
			R_{\mathrm{tel}} = \frac{(V_{\mathrm{dark}} - M_{\mathrm{inst}}R_{\mathrm{inst}})}{M_{\mathrm{tel}}}\hspace{0.5cm}\displaystyle{\left( \frac{\mathrm{V}\,\mathrm{GHz^{-1}}}{\mathrm{W}\ \mathrm{m^{-2}}\ \mathrm{sr^{-1}}\ \mathrm{Hz^{-1}}}\right)}
		\end{equation}

		\noindent
		where \rtel~is referred to as the telescope {\it relative spectral response function}~(RSRF), although it also contains the absolute conversion from \vghz~to surface brightness in units of flux density per steradian. \surfb. $V_{\mathrm{dark}}$, which has the unit of $\mathrm{V}\,\mathrm{GHz^{-1}}$, is the voltage density when observing the dark sky. $M_{\mathrm{inst}}$ and $M_{\mathrm{tel}}$, which have the unit of $\mathrm{W}\ \mathrm{m^{-2}}\ \mathrm{sr^{-1}}\ \mathrm{Hz^{-1}}$, are the intensities, calculated from the measured temperatures, of the instrument and the telescope. $R_{\mathrm{inst}}$ is the instrument RSRF (the spectral response to the instrument is different to that of the main beam). The observed source voltage density, $V_{\mathrm{obs}}$, is then converted to surface brightness, $I_{\mathrm{ext}}$, using,

		% Equation 2
		\begin{equation}
			\label{eq1b}
			I_{\mathrm{ext}} = \frac{(V_{\mathrm{obs}} - M_{\mathrm{inst}}R_{\mathrm{inst}})}{R_{\mathrm{tel}}} - M_{\mathrm{tel}}\hspace{0.5cm}(\mathrm{W}\ \mathrm{m^{-2}}\ \mathrm{sr^{-1}}\ \mathrm{Hz^{-1}})
		\end{equation}
                }

		% Section 3.1, Paragraph 2
		\par{This calibration is only suitable for a uniformly extended source, which means that the solid angle of the source light distribution profile is assumed to be much larger than the that of the FTS beam at any given frequency, $\nu$, i.e. $\Omega_{\mathrm{source}}(\nu)\gg\Omega_{\mathrm{beam}}(\nu)$. For point-like sources, the Level-1 data are not meaningful, but act as an intermediate step to the Level-2 data.
		}

	% Section 3.2
	\subsection{Level-2: Point-Source Calibration}\label{subsec_ps_calibration}

		% Sectuib 3.2, Paragraph 1
		\indent\par{
		In the second step of the pipeline, $I_{\mathrm{ext}}$ is converted into flux density, in units of Jy, with a point-source calibration based on observations of Uranus. This calibration uses a point-source conversion factor, $C_{\mathrm{point}}$,

		% Equation 3
		\begin{equation}
			\label{eq2a}
			C_{\mathrm{point}} = \frac{M_{\mathrm{uranus}}}{I_{\mathrm{uranus}}}\hspace{0.5cm}(\mathrm{sr}),
		\end{equation}

		% Equation 4
		\begin{equation}
			\label{eq2b}
			F_{\rm point} = I_{\rm ext}\ C_{\rm point}\hspace{0.5cm}(\mathrm{Jy})
		\end{equation}

		\noindent where $M_{\mathrm{uranus}}$ is the modeled flux density spectrum of Uranus as described in \citet{orton}, and $I_{\mathrm{uranus}}$ is the Level-1 data from the observation of Uranus. The model is corrected to take into account the finite size of the planet on the day of observation, giving a true point source calibration (assuming good telescope pointing). The resulting flux density (in $\mathrm{Jy}$), $F_{\rm point}$, forms Level-2 data in the pipeline. This calibration is only suitable when the observed source is point-like, i.e. $\Omega_{\mathrm{source}}(\nu)\ll\Omega_{\mathrm{beam}}(\nu)$.
		}

% Section 4
\section{Correction for the Semi-extended Source Distribution}\label{sec_semi}

	% Section 4, Paragraph 1
	\indent\par{
	While the standard pipeline process for analyzing SPIRE FTS data accommodates either point-like or fully extended sources (see Section~\ref{sec_calibration}), in practice, most astronomical sources fall between these two extreme cases. The effects of improperly applied calibration can become pronounced in the overlap region of the two bands of SPIRE FTS because the SLW beam diameter is a factor of $\sim2$ larger than that of SSW (see Figure~\ref{fig_beam_modes}). Figure~\ref{fig_point_ext} illustrates the resulting spectra of point and extended calibrated data for sources of increasing angular size (point-like, semi-extended and fully extended). When the calibration is inappropriate for the extent of the source, a noticeable difference in the measured flux and slope of the spectrum is observed at the overlapping region of the two bands. This is caused by the abrupt change in the effective beam size at those frequencies (see Section~\ref{sec_beam}). For example, Figure~\ref{fig_point_ext}-(f) shows the case in which a single-pointed, sparsely sampled observation of a uniformly extended source has been calibrated as a point source. It can be clearly seen that the resulting spectra show a distorted shape with a large step between the two bands, in which the SSW spectrum is underestimated. By comparison, the extended-source calibrated spectra are smooth and have no step between the bands, as shown in Figure~\ref{fig_point_ext}-(c). Similarly, the spectrum of a true point source would appear smooth when point-source calibrated (e.g. Figure~\ref{fig_point_ext}-(d)), but distorted when extended-source calibrated (e.g. Figure~\ref{fig_point_ext}-(a)).
	}

	% Section 4, Paragraph 2
	\par{
	An appropriately calibrated spectrum should contain no discontinuity between the SSW and SLW bands. Thus, the difference in intensity in the overlap region of the two bands, in both point-source and extended-source calibrated data, provides a simple method of estimating the source intensity distribution. This is particularly important for sparsely sampled SPIRE FTS observations, because in the absence of corresponding higher spatial resolution maps, the intensity jump  may be the only indication of angular extent of the observed object.
	}

	% Section 4, Paragraph 3
	\par{
	The remainder of this section describes a method for correcting a SPIRE FTS spectrum based on an assumed source intensity distribution.
	}
	
	% Section 4.1
	\subsection{Correction for the Source-Distribution}
	\label{correction}
	
		% Figure 3: Signal through telescope diagram
		\begin{figure}
			\centering
			\includegraphics[width=0.5\textwidth]{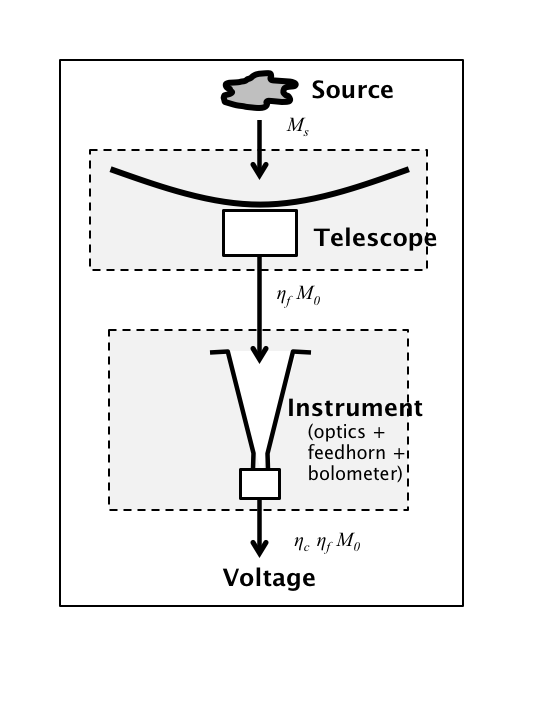}\vspace{-1cm}
			\caption{A diagram demonstrating how signals from a source, $M_{\mathrm{s}}$, change before reaching the bolometer.}
			\label{source_telescope}
		\end{figure}

		% Section 4.1, Paragraph 1
		\indent\par{
		The goal of calibrating the FTS observation of a semi-extended source is to recover the total flux density~(in Jy) from the source signal that has reached the bolometers. In order to achieve this, a conversion factor for a given source should be calculated, similar to Equation~\ref{eq2a}, as follows,

		% Equation 5
		\begin{equation}
			C_{\mathrm{s}}=\frac{M_{\mathrm{s}}}{I_{\mathrm{ext}}}\hspace{0.5cm}(\mathrm{sr})
			\label{cs}
		\end{equation}

		\noindent
		where $M_{\mathrm{s}}$ is the true integrated {\it flux density} of the source. Once $C_{\mathrm{s}}$ has been determined, it is possible to obtain the flux density by multiplying $I_{\mathrm{ext}}$~(from Equation~\ref{eq1b}) by $C_{\mathrm{s}}$, or by multiplying the point-source calibrated data, $F_{\mathrm{point}}$~(from Equation~\ref{eq2b}), by a factor $\displaystyle{\frac{C_{\mathrm{s}}}{C_{\mathrm{point}}}}$.
		}

		% Section 4.1, Paragraph 2
		\par{
		$C_{\mathrm{s}}$ is impossible to measure, because determining the true flux density, $M_{\mathrm{s}}$, is, after all, the purpose of the observation that is being calibrated. However, $C_{\mathrm{s}}$ can be constructed by assuming a dimensionless source light distribution, $D_{\nu}(\Psi)$, and an understanding of the beam profile, $P_{\nu}(\Psi)$. $D_{\nu}(\Psi)$ and $P_{\nu}(\Psi)$ define the solid angles of the source, $\Omega_{\mathrm{source}}$, and of the beam, $\Omega_{\mathrm{beam}}$, with the following relationship,
		
		% Equation 6
		\begin{equation}
			\centering
			\label{source_area}
			\Omega_{\mathrm{source}}(\nu) = \displaystyle\iint\limits_{2\pi}\,D_{\nu}(\Psi)\,d\Psi\hspace{0.5cm}(\mathrm{sr})
		\end{equation}

		% Equation 7
		\begin{equation}
			\label{beam_area}
			\Omega_{\mathrm{beam}}(\nu) = \displaystyle\iint\limits_{2\pi}\,P_{\nu}(\Psi)\,d\Psi\hspace{0.5cm}(\mathrm{sr})
		\end{equation}

		\noindent
		If $D_{\nu}(\Psi)$ provides a good representation of the source spatial profile, a relationship between $M_{\mathrm{s}}$ and $D_{\nu}(\Psi)$ can be established,
		
		% Equation 8
		\begin{equation}
			M_{\mathrm{s}}=M_{0}\,\iint\limits_{beam}\,D_{\nu}(\Psi)\,d\Psi=M_{0}\,\Omega_{\mathrm{source}}(\nu)\hspace{0.5cm}(\mathrm{Jy}),
			\label{ms}
		\end{equation}
				
		\noindent
		where $M_{0}$ is a measure of the average {\it surface brightness} of the source and has the unit of \surfb.
                }

                % Section 4.1, Paragraph 3
                \par{
                The propagation of the source model spectrum through the telescope and instrument is summarized in Figure~\ref{source_telescope}. The efficiency with which the source couples to the telescope is defined by the {\it forward coupling efficiency}~(see \citealt{ulich-apjs-1976, kutner81}), $\eta_{\mathrm{f}}$, 
		
		% Equation 9
		\begin{equation}
			\eta_{\mathrm{f}}(\nu,\Omega_{\mathrm{source}}) = \frac{\displaystyle\iint\limits_{2\pi}\ P_{\nu}(\Psi-\Omega_{0})\,D_{\nu}(\Psi)\,\mathrm{d}\Psi}{\displaystyle\iint\limits_{2\pi}P_{\nu}(\Psi)\,\mathrm{d}\Psi}
			\label{etaf}
		\end{equation}
		
		\noindent
		where $\Omega_{0}$ accounts for any offset of the source from the center of the beam (the position of the source inside the beam clearly affects how the source distribution couples to the beam profile).}

                % Section 4.1, Paragraph 4
		\par{
		We introduce an empirically derived factor, the correction efficiency~($\eta_{\mathrm{c}}$), to deal with additional effects that are not already taken into account by Equation~\ref{etaf}. Firstly, the efficiency with which the reconstructed beam profile is coupled to the source distribution is not known. Secondly, the efficiency with which Equation~\ref{ms} represents the true source is not known, and finally the beam profile was only measured over a finite solid angle and so the reconstruction does not take account of the response far from the optical axis. For a point source, we expect $\eta_{\mathrm{c}}(\nu,\Omega_{\mathrm{point}})=1$, i.e. the measured beam profile, $P_{\nu}(\Psi)$, is sufficient to represent the effective solid angle covered by the source. For semi-extended sources, in principle, $\eta_{\mathrm{c}}(\nu,\Omega_{\mathrm{source}})$ could be determined by decomposing the source distribution using the same technique discussed in Section~\ref{sec_beam}. For computational efficiency, we only treat the total source and beam separately in this work. As presented in Equation~\ref{eq1b}, $I_{\rm ext}$ is the observed voltage density~(see Figure~\ref{source_telescope}) calibrated with observations of dark sky. That is, instead of applying an $\eta_{c}$ proper for the observed source, an $\eta_{c}$ suitable for an observation of dark sky is applied to $I_{\mathrm{ext}}$. With the $\eta_{\mathrm{c}}$ factor taken into account, $I_{\mathrm{ext}}$ can be formulated as,

		% Equation 10
		\begin{align}
			I_{\mathrm{ext}}=M_{0}\,\frac{\eta_{\mathrm{c}}(\nu,\Omega_{\mathrm{beam}})}{\eta_{\mathrm{c}}(\nu,\Omega_{\mathrm{source}})}\,\eta_{\mathrm{f}}(\nu,\Omega_{\mathrm{source}})\hspace{0.5cm}(\mathrm{W}\ \mathrm{m^{-2}}\ \mathrm{sr^{-1}}\ \mathrm{Hz^{-1}}).
			\label{isource}
		\end{align}
		
		\noindent
		Equation~\ref{cs} can then be re-written as,

		% Equation 11
		\begin{equation}
			C_{\mathrm{s}}=\displaystyle{\frac{\eta_{\mathrm{c}}(\nu,\Omega_{\mathrm{source}})}{\eta_{\mathrm{c}}(\nu,\Omega_{\mathrm{beam}})}\,\frac{\Omega_{\mathrm{source}}}{\eta_{\mathrm{f}}(\nu,\Omega_{\mathrm{source}})}}\hspace{0.5cm}(\mathrm{sr}).
			\label{cs_form}
		\end{equation}
		}

		% Section 4.1, Paragraph 5
		\par{
		In the two extreme cases, i.e. when the source is uniformly-extended or point-like, Equation~\ref{cs_form} should be consistent with the calibration data from the FTS pipeline:
		}

		% Section 4.1, case I
		\paragraph{Case~\textsc{I}: $\Omega_{\mathrm{source}}(\nu)\gg\Omega_{\mathrm{beam}}(\nu)$}\mbox{}\\
		
			\noindent
			In this extreme case, the source can be locally regarded as uniformly extended at the scale of the FTS beam, so $D_{\nu}(\Psi)=1$. This means that $\eta_{\mathrm{c}}(\nu,\Omega_{\mathrm{source}})=\eta_{\mathrm{c}}(\nu,\Omega_{\mathrm{beam}})$, and $\eta_{\mathrm{f}}$ is reduced to 1. If one would like to recover the total flux of the source, Equation~\ref{cs_form} can be expressed as,
			
			% Equation 12
			\begin{equation}
				C_{\mathrm{ext}} =\Omega_{\mathrm{source}}\hspace{0.5cm}(\mathrm{sr})
				\label{cext}
			\end{equation}
			
			\noindent
			In Equation~\ref{cext}, $\Omega_{\mathrm{source}}$ for a uniformly extended source will be determined by the reference solid angle~($\Omega_{\mathrm{beam}}$), within which the source can be regarded as uniformly extended. 

		% Section 4.1, case II
		\paragraph{Case~\textsc{II}: $\Omega_{\mathrm{source}}(\nu)\ll\Omega_{\mathrm{beam}}(\nu)$}\mbox{}\\
			
			\noindent
			In the case of a point source, $\Omega_{\mathrm{source}}\sim\delta\Omega\ll\Omega_{\mathrm{beam}}$. Since the beam profile has a value of unity at the center, $\displaystyle{\eta_{\mathrm{f}}(\nu)\simeq\frac{\delta\Omega}{\Omega_{\mathrm{beam}}(\nu)}}$. Based on Equation~\ref{cs_form}, $C_{\mathrm{point}}$ can be written as
			
			%Equation 13
			\begin{equation}
				C_{\mathrm{point}}=\displaystyle{\frac{\Omega_{\mathrm{beam}}(\nu)}{\eta_{\mathrm{c}}\,(\nu,\Omega_{\mathrm{beam}})}}\hspace{0.5cm}(\mathrm{sr})
				\label{cpoint}
			\end{equation}

			\noindent
			where, as previously mentioned, $\eta_{\mathrm{c}}$ for a point source has been set to unity in Equation~\ref{cpoint}.
			\vspace{18pt}
		
		% Section 4.1, Paragraph 4
		\par{
		The final correction for intermediate cases, based on Equation~\ref{cs_form}, can be applied to either the extended-source calibrated~($I_{\mathrm{ext}}$) or point-source calibrated~($F_{\mathrm{point}}$) data with the following equations,

		% Equation 14
		\begin{equation}
			\label{source_correction}
			F_{\mathrm{s}}=
			\begin{cases}
				\displaystyle{\,I_{\mathrm{ext}}\cdot\, \frac{\eta_{\mathrm{c}}(\nu,\Omega_{\mathrm{source}})}{\eta_{\mathrm{c}}(\nu,\Omega_{\mathrm{beam}})}\,\frac{\Omega_{\mathrm{source}}}{\eta_{\mathrm{f}}(\nu,\Omega_{\mathrm{source}})}}\hspace{0.5cm}(\mathrm{Jy}) \\ \\
				\displaystyle{\,F_{\mathrm{point}}\cdot\eta_{\mathrm{c}}(\nu,\Omega_{\mathrm{source}})\,\frac{\Omega_{\mathrm{source}}}
				{\eta_{\mathrm{f}}(\nu,\Omega_{\mathrm{source}})\,\Omega_{\mathrm{beam}}(\nu)}}\hspace{0.5cm}(\mathrm{Jy})
			\end{cases}
		\end{equation}
		}
		
		% Section 4.1, Paragraph 5
		\par{For most astronomical objects, the light distribution changes little in the bandwidth of the FTS as seen in the SPIRE photometry measurements. Therefore, for the purpose of the correction, $D_{\nu}(\Psi)$ is assumed to be independent of frequency. This general assumption may not be suitable for the distribution of the spectral lines that often originate in regions of the source with spatial extents that can vary greatly from line to line, from species to species, and from line to continuum. The correction derived in this section, as given in Equation~\ref{source_correction}, gives the total flux density from the source as defined by $D_{\nu}(\Psi)$. If this covers a large solid angle, it may actually be desirable to limit the flux density to that which is within a reference beam. In order to define a consistent spectrum, the reference beam should be constant in frequency across the band. It could, for example, be set to a Gaussian that approximates the beam of another telescope. The effect of a reference beam modifies the solid angle of the source in the numerator of Equation~\ref{source_correction} to be,

		% Equation 15
		\begin{equation}
			\label{reference_beam}
                        \Omega_{\mathrm{source-ref}} = \displaystyle\iint\limits_{2\pi}\ P_{\mathrm{ref}}(\Psi)\,D_{\nu}(\Psi)\,\mathrm{d}\Psi\hspace{0.5cm}(\mathrm{sr})
		\end{equation}
		}

	% Section 4.2
	\subsection{Determination of the Correction Efficiency}\label{subsec_etaem}

		% Section 4.2, Paragraph 1
		\indent\par{
		To understand how $\eta_{\mathrm{c}}$ is affected by different source sizes, observations of several planets, with well defined source size and spectral flux density, were examined. Table~\ref{sect_source} presents the sources and observations used. Assuming their light distribution is well described by a circular top-hat model, the model spectrum for each planet was divided by the source-distribution corrected spectrum from Equation~\ref{source_correction} to calculate $\eta_{\mathrm{c}}$ as a function of frequency,

		% Equation 16
		\begin{equation}
			\displaystyle{\eta_{\mathrm{c}}(\nu,\Omega_{\mathrm{source}})\,=\,\frac{F_{\mathrm{model}}}{F_{\mathrm{point}}}\,\frac{\eta_{\mathrm{f}}(\nu,\Omega_{\mathrm{source}})\,\Omega_{\mathrm{beam}}(\nu)}{\Omega_{\mathrm{source}}}}.
		\end{equation}

		% Table 1
		\begin{table}
			\caption{Properties of the sources used to derive the correction efficiency.}
			\label{sect_source}
			\begin{minipage*}{8.8cm}
				\small
				\begin{center}
					\begin{tabular}{|c c c c c|}
						\hline\hline
                                                Observation ID & Date & OD~\footnotemark[1] & Object & Diameter (")~\footnotemark[2]\\
						\hline       
                                                1342221703 & 2011-05-26 & 742 & Neptune &  2.3 \\            
                                                1342257307 & 2012-12-16 & 1313 & Uranus & 3.6\\
                                                1342197462 & 2011-10-16 & 383 & Mars & 6.0\\
                                                1342224754 &  2011-07-25 & 803 & Saturn & 16.7\\
					\hline\hline                  
					\end{tabular}
					\footnotetext[1]{
					\textsl{Herschel} Operational Day.
					}
                                        \footnotetext[2]{
					Simulated with the NASA JPL Solar System Simulator
					}
				\end{center}
			\end{minipage*}
		\end{table}

		% Figure 4: Eta_c function
		\begin{figure}
			\centering
			\includegraphics[width=\hsize]{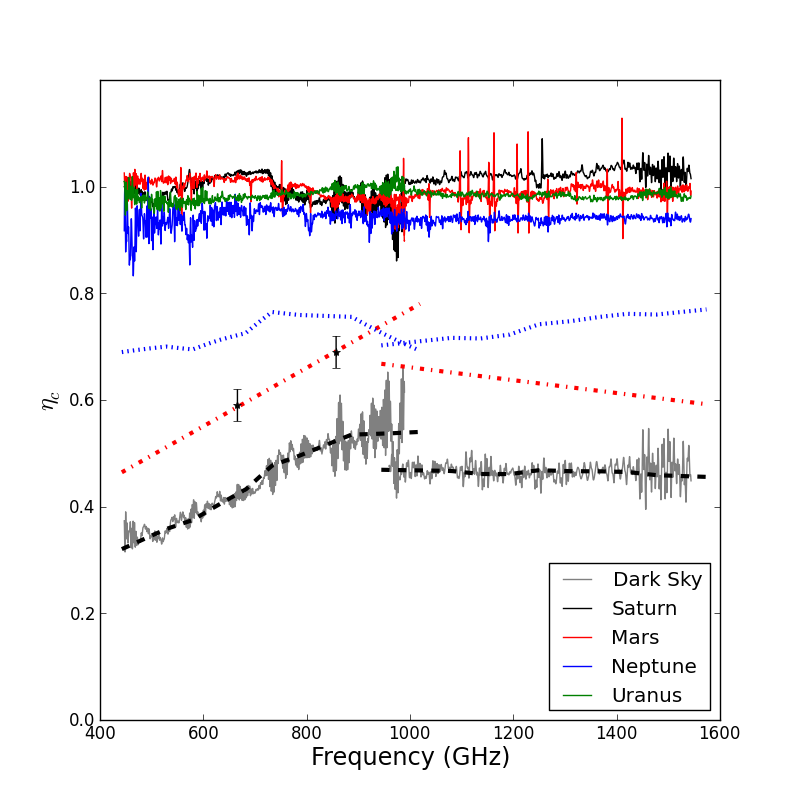}
			\caption{The correction efficiency $\eta_{\mathrm{c}}(\nu,\Omega_{\mathrm{source}})$ for the sources listed in Table~\ref{sect_source} and dark sky~(grey). The plot shows the variation of $\eta_{\mathrm{c}}$ with the frequency for each source. The blue dotted lines indicate the efficiency predicted by the optics model. The black dashed lines indicate a smoothed efficiency function of an extended source. The ratio between the two is indicated by the red dash dot lines. The two black stars are the far field efficiency measured from prototype arrays in the lab~\citep{chattopadhyay-ieeetrans-2003}.}
			\label{eta}
		\end{figure}

		\noindent
		The derivation of the model of Saturn is described in detail in \citet{fletcher}. This model was scaled to the observation of Saturn on 2011-07-25~(OD 803) based on the distance to the planet at the time of the observation. The contribution of Saturn's rings was found to be negligible ($<1\%$) for the data observed in 2010 (ObsID 1342198279,~\citealt{fletcher}). On the day of the Saturn observation used in this work (ObsID 1342224754), the planet phase angle, as simulated by the NASA JPL Solar System Simulator~\footnote{http://space.jpl.nasa.gov/}, is only 0.4$^{\rm\circ}$ smaller, and the position angle of the rings is slightly larger than in 2010. However, within the other uncertainties of the measurement, the contribution of the rings can still be considered as negligible. The models of Uranus and Neptune are described in \citet{swinyard-aa-2013}. A high spectral resolution model of Mars was constructed based on the thermophysical model of \citet{rudy}, updated to use the thermal inertia and albedo maps (0.125 degree resolution) derived from the Mars Global Surveyor Thermal Emission Spectrometer (5.1--150 $\mu$m) observations \citep{putzig}. These new maps were binned to 1 degree resolution. A dielectric constant of 2.25 was used for latitudes between 60 degrees South and 60 degrees North. As in the original Rudy model, surface absorption was ignored in the polar regions and a dielectric constant of 1.5 in the CO$_{2}$ frost layer was assumed. Disk-averaged brightness temperatures were computed over the SPIRE frequency range and converted to flux densities for the times of the observations. This model was then photometrically matched to the continuum model of Lellouch (available on the internet~\footnote{http://www.lesia.obspm.fr/perso/emmanuel-lellouch/mars/}) for the time of the Mars observation on 2011-10-16~(OD 383).
		}

		% Section 4.2, Paragraph 2
		\par{
		Figure~\ref{eta} shows $\eta_{\mathrm{c}}$ as a function of frequency for the listed sources. The average values of $\eta_{\mathrm{c}}$ for Neptune, Uranus, Mars, and, Saturn, are 0.94, 0.98, 0.98, and, 1.02, respectively for SLWC3, and, 0.94, 0.98, 0.99, and, 0.99, respectively for SSWD4. The results indicate that for the sources with angular size up to 16.7"~(the size of Saturn on OD803), which is slightly larger than the smallest FWHM, 16.55'' of the FTS beam, the correction efficiency is nearly 1. Due to the difficulty of selecting well--studied sources with angular sizes between Saturn and an extended source, Figure~\ref{eta} also contains the results for a fully extended source (e.g. an observation of dark sky). In this case, $F_{\mathrm{model}}$ was calculated following Equation~\ref{cext}, such that $\eta_{\mathrm{c}}=\Omega_{\mathrm{beam}}/C_{\mathrm{point}}$ - i.e. the data from the observation itself cancels and the plot shows the limiting case for a fully extended source (grey curve in Figure~\ref{eta}). This result shows that the correction works at least on the sources that are as large as or smaller than the smallest FWHM of the SPIRE FTS.
		}
		
		% Section 4.2, Paragraph 3
		\par{
		In Figure~\ref{eta}, the apparent difference indicated by the correction efficiency between coupling to an extended source (grey curve) and a point like source~(blue and green curves) illustrated here has several possible causes. The most obvious one is loss of flux due to diffraction of the beam as it passes through the optics of the telescope and instrument. Whilst the number of optical elements within the system precludes developing a detailed physical model, the diffraction loss versus frequency for a point source on axis can be obtained from a simple physical optics model using Fourier transforms. The pupil mask within SPIRE is designed to limit the detector spatial response with an edge taper of $8\,\mathrm{dB}$, and there is additional obscuration from the secondary mirror and the secondary mirror supports~\citep{caldwell2000}. A simple representation of this system was used to make a first order Fourier transform optical model. The results of the model are shown as blue dotted lines in Figure~\ref{eta} together with $\eta_{\mathrm{c}}$ for an extended source (gray line), and the residual~(red dash dot lines), which is the ratio between a smoothed version of $\eta_{\mathrm{c}}$ for an extended source (black dashed lines) and the diffraction model results. The residual is seen to be essentially linear with the detailed shape of the efficiency curve replicated by the diffraction losses.  This suggests that diffraction is responsible for part of the difference in coupling efficiency and some other factor accounts for the residual.  One plausible explanation for the residual losses is a difference in the coupling efficiency of a point and extended sources within the feedhorns and bolometers. This is supported by comparing the residual linear efficiency curve to the results of the measurement of the far field coupling efficiency of the feedhorns and bolometers themselves (indicated by the black stars in Figure~\ref{eta}).  These measurements were taken in the absence of any fore optics and are reported in \citet{chattopadhyay-ieeetrans-2003}. The dependence on frequency looks similar between the residual and the~\citet{chattopadhyay-ieeetrans-2003} results indicating that the possible cause of the apparently low correction efficiency for extended sources is due to a combination of diffraction losses and a difference in the response of the feed horns and bolometers to a source fully filling the aperture and that to a point source.
		}
				
	% Section 4.3
	\subsection{Determine Source Size Using Saturn}\label{subsec_srcsize}

		% Figure 5: Saturn S/N analysis
		\begin{figure}
			\centering
			\includegraphics[width=\hsize]{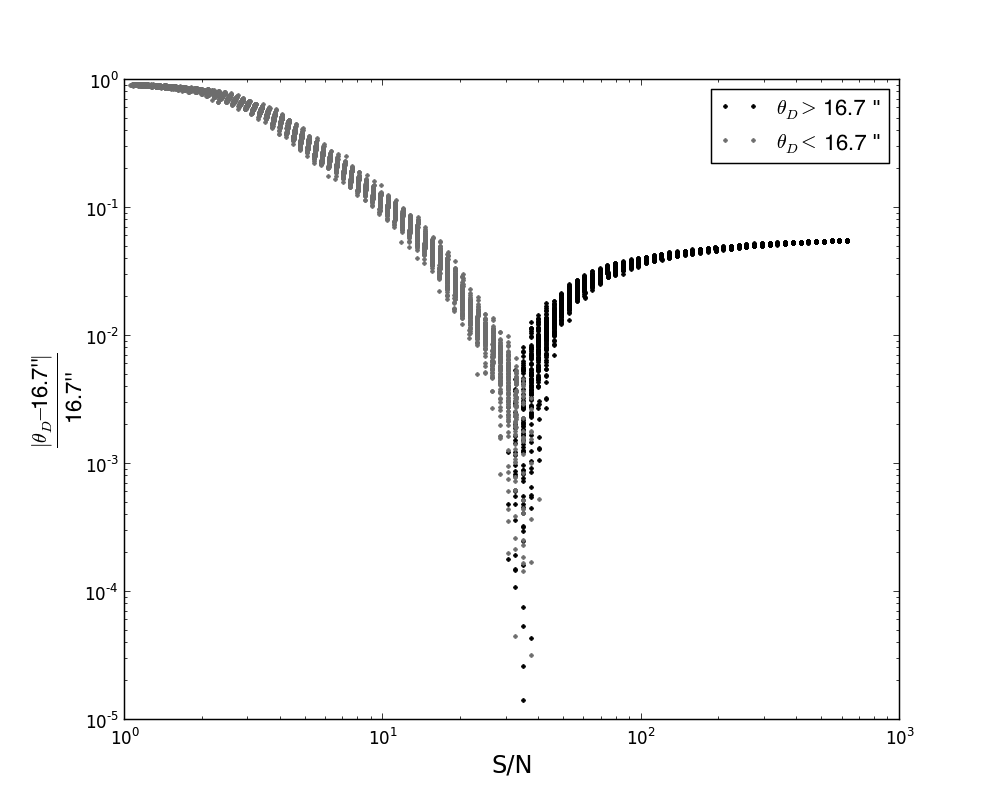}
			\caption{A diagram showing how the best-fit $\theta_{\mathrm{D}}$ varies with $S/N$ for Saturn. The grey color indicates the instances where the best-fit $\theta_{\mathrm{D}}$ is below 16.7'', which is the true angular size of Saturn for the observation. The black color indicates where the best-fit $\theta_{\mathrm{D}}$ is above 16.7''.}
			\label{saturn_sn}
		\end{figure}

		% Figure 6: Corrected planets spectra
		\begin{figure}
			\centering
			\includegraphics[height=7.5cm]{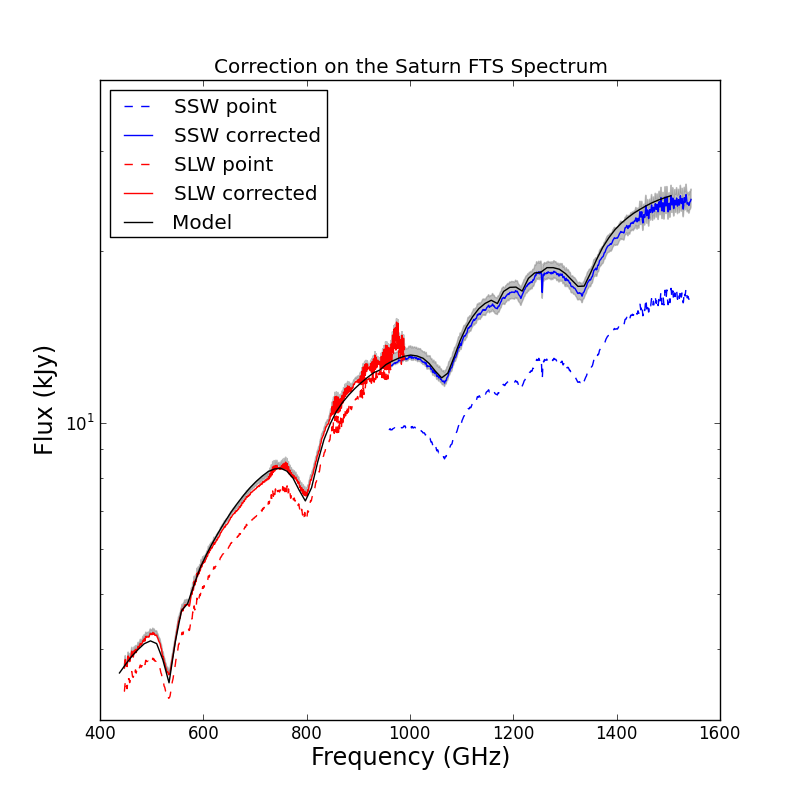}
			\caption{The spectrum of Saturn corrected by a top-hat disk model with 17.2'' in its angular size. The gray color indicates the values between two spectra corrected with a Gaussian profile with $\theta_{\mathrm{D}}=12''$ and a Sersic profile with $\theta_{\mathrm{D}}=4''$.}
			\label{planet_corrected}
		\end{figure}
		
		% Figure 7: Overlap Bandwidth
		\begin{figure}
			\centering
			\includegraphics[width=\hsize]{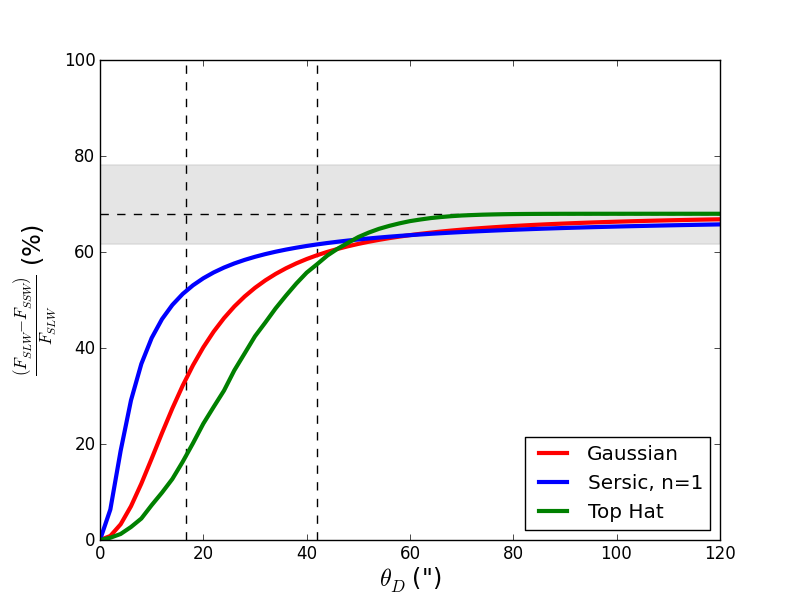}
			\caption{Demonstration of how the angular size of source affects the flux difference at the overlap bandwidth for a top-hat (green), a Gaussian (red), and, a Sersic $n=1$ (blue) profile. The two vertical dashed lines indicate the largest and the smallest beam FWHM of the FTS at 42'' and 16.55''. The horizontal dashed line indicates the percentage difference of the fluxes if a source uniformly fills the beams without taking $\eta_{\mathrm{c}}$ in account, while the grey band indicates the same value but accounts for the uncertainty introduced by $\eta_{\mathrm{c}}$.}
			\label{overlap_size}
		\end{figure}

		% Section 4.3, Paragraph 1
		\indent\par{
		The SLW and SSW bands overlap in the region of $959<\nu<989$~GHz. As already shown in Figure~\ref{fig_beam_modes}, the FWHM of SLW in this frequency range is approximately twice that of SSW. If the correction described by Equation~\ref{source_correction} is applied using the fitted SPIRE profiles \citep{makiwa-prep-2013}, the spectra from SLW and SSW should match in the overlap bandwidth. The Saturn observation, of which the true intensity distribution is well known, can be used to estimate the sensitivity of the correction to source size, to give an indication of the uncertainties in the method. The diameter of Saturn's disk on OD803 was 16.7'' and the observed spectrum has an averaged signal-to-noise ratio~($S/N$) of $\sim630$ in the overlapping frequency range. This makes it a perfect reference case with a well-constrained source distribution and high $S/N$.
		}

		% Section 4.3, Paragraph 2
		\par{
		The error of estimation was propagated based on a series of simulated observations generated from the observed fluxes~($f_{\mathrm{o}}$) and errors~($\epsilon_{\mathrm{o}}$) using a Monte-Carlo method. For each simulated observation, the observed errors were multiplied by a fixed number which varied between 1 and 600 to generate simulated errors~($\epsilon_{\mathrm{s}}$). The simulated flux density~($f_{\mathrm{s}}$) was then {\it randomly} generated to be in the range, $(f_{\mathrm{o}}-\epsilon_{\mathrm{s}})<f_{\mathrm{s}}<(f_{\mathrm{o}}+\epsilon_{\mathrm{s}})$. For each simulated observation, a 2-D circular top--hat model of the planet size was created with a range of diameters, $3''<\theta_{\mathrm{D}}<23''$. These were then used in Equation~\ref{source_correction} and the best-fit angular size selected by minimizing the $\chi^{2}$ parameter for the overlap region,

			% Equation 17:
			\begin{equation}
				\chi^{2}(\theta_{\mathrm{D}})=\displaystyle{\sum_{i}\frac{F_{\mathrm{SLW}}(\nu_{i},\theta_{\mathrm{D}})-F_{\mathrm{SSW}}(\nu_{i},\theta_{\mathrm{D}})}{\sigma_{\mathrm{SLW}}(\nu_{i})^{2}+\sigma_{\mathrm{SSW}}(\nu_{i})^{2}}}
				\label{size_chisq}
			\end{equation}

		\noindent where $F_{\mathrm{SLW}}(\nu_{i},\theta_{\mathrm{D}})$ and $F_{\mathrm{SSW}}(\nu_{i},\theta_{\mathrm{D}})$ are the flux densities of Saturn corrected with Equation~\ref{source_correction} on a simulated spectrum; $\sigma_{\mathrm{SLW}}(\nu_{i})$ and $\sigma_{\mathrm{SSW}}(\nu_{i})$ are the simulated errors for SLW and SSW, respectively.
		}
		
		% Section 4.3, Paragraph 3
		\par{
		Considering only the statistical errors, the best-fit $\theta_{\mathrm{D}}$ is 17.2'', which is about 3\% larger than the true value of 16.7'', and the reduced $\chi^{2}$ is 60.62. The large value of reduced $\chi^{2}$ reflects that the statistical errors under-represent the total errors of the spectrum in Equation~\ref{size_chisq}. Figure~\ref{saturn_sn} shows how the best-fit $\theta_{\mathrm{D}}$ varies with the simulated $S/N$. The best-fit $\theta_{\mathrm{D}}$ is systematically smaller than the true size when $S/N<35$~(indicated with grey color) and converges to 17.2'' with increasing $S/N$ above $S/N=35$~(indicated with black color). The best-fit $\theta_{\mathrm{D}}$ is closest to 16.7'' at the simulated $S/N\sim 35$ with a reduced $\chi^{2}\sim0.18$. At $S/N\sim80$, the reduced $\chi^{2}=1.01$ for $\theta_{\mathrm{D}}=17.0$. This means that beyond $S/N\sim100$, where the best-fit $\theta_{\mathrm{D}}$ starts to converge, increase of $S/N$ does not affect the estimated source size, and the estimation is only applicable when the observation has at least $S/N\sim30$. Figure~\ref{planet_corrected} shows the FTS spectrum of Saturn corrected by a top-hat intensity distribution with $\theta_{\mathrm{D}}=17.2''$, the best fit found with the least $\chi^{2}$ analysis at the overlap bandwidth. With consideration of only the source distribution ($\eta_{\mathrm{c}}=1$ has been adopted throughout the derivation) the distribution--corrected spectrum shows good agreement with the model spectrum.
		}
		
		% Section 4.3, Paragraph 4
		\par{
		The above analysis indicates that the determined angular size has $\lesssim3\%$ deviation from the true size when the observed spectrum has $S/N>10$. The large reduced $\chi^{2}$ implies that there are errors that were unaccounted for in the analysis. Based on the fact that the reduced $\chi^{2}=1.01$ at $S/N\sim80$, the total error of observation should be $\sim1.2\%$, instead of $\sim0.1\%$ as originally indicated by the $S/N$ of the observation. The increase of error can be due to under-estimated errors from the pipeline and/or the over-simplified model for the light distribution of Saturn. It should also be noted that the above derivation assumes a Gaussian distribution of the errors.
		}

		% Figure 8: M82		
		\begin{figure*}
			\centering
			%Figure 8a: M82 spectrum
			\begin{subfigure}{0.45\textwidth}
				\includegraphics[width=7.5cm]{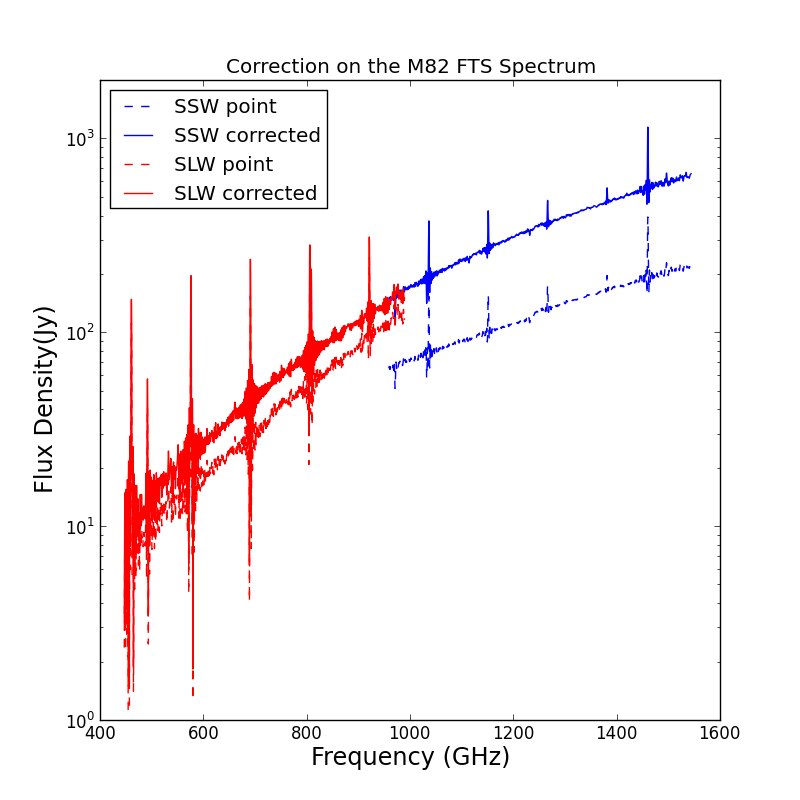}
				\caption{The spectrum of M82 observed by the FTS on 2010-11-08~(OD 543). The red and blue colors indicate the spectra observed by the SLWC3 and SSWD4, respectively. The dashed line shows the spectrum calibrated by the point-source calibration mode. The solid line shows the spectrum corrected by the method presented in this work with a best-fit source $r_{\mathrm{e}}=11''$~(see Equation~\ref{gal_profile}.)}
				\label{m82_spec}
			\end{subfigure}
			\quad
			%Figure 8a: M82 CO SLED
			\begin{subfigure}{0.45\textwidth}
				\includegraphics[width=7.5cm]{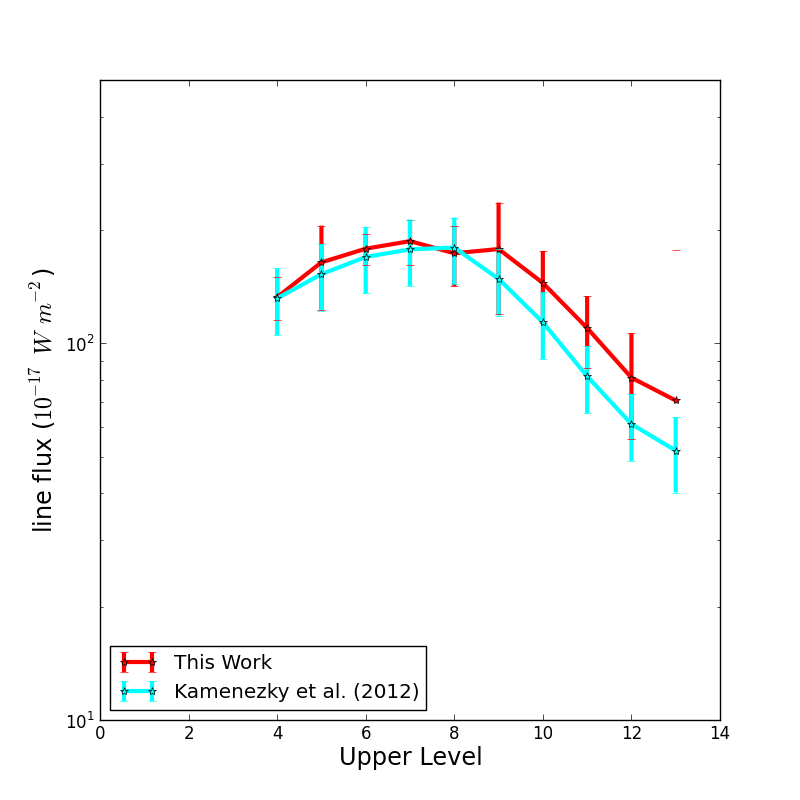}
				\caption{A comparison of the CO SLEDs from the starburst core of M82. The red color indicates the fluxes derived from the central detector spectrum (SLWC3 and SSWD4), corrected for the source distribution and convolved to a Gaussian beam of FWHM$=42''$~(red). The cyan color is the same spectrum corrected by a source-beam coupling factor derived from the SPIRE photometry $250\,\mum$ map with appropriate profiles to produce the continuum light distribution seen with the FTS~(cyan, \citealt{kamenezky-apj-2012}).}
				\label{m82_sled}
			\end{subfigure}
			\caption{The M82 FTS spectrum corrected with the method developed in this work.}
		\end{figure*}

		% Section 4.3, Paragraph 5
		\par{
		The determination of source size using the overlap bandwidth is carried out by minimizing differences between the fluxes measured in SSW and SLW. Figure~\ref{overlap_size} shows the relationship between the percentage difference in the overlap bandwidth and the angular size of the source light profile (FWHM when applicable). The figure shows profiles described by a top--hat~(green), Sersic profile with $n=1$~(blue), and, Gaussian model~(red), with the naive assumption that $\eta_{\mathrm{c}}=1$ throughout. At large angular size, the percentage difference approaches 68\%~(indicated by the horizontal dashed line), which can be well explained by the ratio of the beam solid angles of SSW and SLW in the overlap bandwidth region~(approximately a ratio of 0.32). The horizontal grey shaded band in Figure~\ref{overlap_size} shows the effect of taking $\eta_{\mathrm{c}}$ for a fully extended source (the grey curve in Figure~\ref{eta}) into account in the overlap region. The flux percentage differences of the three example profiles intersect with the grey band at sizes approximately equal to or slightly larger than the largest FWHM of the SLW beam (grey vertical line). The intersect values of 42''~(Sersic profile, $n=1$), 48''~(top-hat profile), and, 50''~(Gaussian profile) mark the limit beyond which a source should be considered an extended source. {\it Based on this result, we recommend to adopt $\eta_{\mathrm{c}}=1$ for sources with a size that is between the FWHM of the SSW and SLW beams in the overlap bandwidth region~($20''<\theta_{\mathrm{D}}<42''$), but that caution should be exercised when interpreting the results with real observations in this range.}}

		% Section 4.3, Paragraph 6
		\par{
		In the case of the Saturn observation used in Figure~\ref{planet_corrected}, the median percentage difference in the overlap bandwidth is $\sim26.3\%$. This value corresponds to a $\theta_{\mathrm{D}}$ of 12'' for a Gaussian profile and of 4'' for a Sersic profile with $n=1$. To examine how the assumption of a source profile affects the corrected spectrum, the same observation of Saturn was corrected with a Gaussian profile with $\theta_{\mathrm{D}}=12''$ and a Sersic profile with $\theta_{\mathrm{D}}=4''$. The gray color in Figure~\ref{planet_corrected} indicates the range between the Gaussian corrected profile (upper boundary) and Sersic corrected profile (lower boundary). The median difference is insignificant~($\sim1\%$) for SLW but rises to $\sim7\%$ for SSW due to the decrease in the beam size. For the SLW beam, either of the assumed profiles is enclosed in the beam, so the difference between the corrected spectra is not significant. For the SSW beam, since the source has a size that is similar to the beam FWHM, the assumed $\theta_{\mathrm{D}}=12''$ Gaussian profile predicts a more extended source than the $\theta_{\mathrm{D}}=17.2''$ top-hat profile. Thus the spectrum is overcorrected to account for the extension of the Gaussian profile. On the other hand, a $\theta_{\mathrm{D}}=4''$ Sersic profile proposes a more compact source which results in a spectrum that is under corrected.
		}

% Section 5
\section{Benchmarking and testing}\label{sec_tests}

	% Section 5, Paragraph 1
	\indent\par{In this section, the correction algorithm derived in Section~\ref{sec_semi} is applied to observations of the nearby galaxy M82 and the compact core of the molecular cloud, Sgr B2. We adopt $\eta_{\rm c}=1$ in this section.}
	
	% Section 5.1
	\subsection{M82}
		% Section 5.1, Paragraph 1
		\indent\par{
		M82 is a nearly edge-on starburst galaxy at 3.4$\,\mathrm{Mpc}$ distance from the Milky-way~\citep{dalcanton-apjs-2009}. The FTS observed M82 in fully sampled mode on the 2010-11-08~(OD 543; ObsID 1342208388). Details of results from the fully sampled map of M82 can be found in \citet{kamenezky-apj-2012}. Here, we apply the correction to the spectrum of M82 taken with the central FTS detectors, SLWC3 and SSWD4, at J2000 coordinates, $(\alpha, \delta)=(9\mathrm{h}55\mathrm{m}52.6\mathrm{s}, +69\mathrm{d}40\mathrm{m}48.02\mathrm{s})$ and compare it to the beam size corrected data taken at the same coordinates with the central detectors~\citep{kamenezky-apj-2012}.
		}
	
		% Section 5.1, Paragraph 2
		\par{
		The signal-to-noise ratio of the FTS spectrum of M82 is $\sim200$. Assuming the intensity distribution at the core of M82 can be described as a 2-D circular exponential profile,
	
			% Equation 18
			\begin{equation}
				B(r)\,=\,exp(-r/r_{e})\hspace{0.5cm}(\mathrm{W}\ \mathrm{m^{-2}}\ \mathrm{sr^{-1}}\ \mathrm{Hz^{-1}})
				\label{gal_profile}
			\end{equation}
		
		\noindent we correct the spectrum of SLWC3 and SSWD4 by matching the fluxes at $959<\nu<989$~GHz with varying scale radius~($8''<r_{e}<14''$).
		}
		
		% Section 5.1, Paragraph 3
		\par{
	        The best-fit $r_{e}$ determined by minimizing $\chi^{2}$, defined in Equation~\ref{size_chisq}, is $\sim11''$, which corresponds to $\theta_{\mathrm{D}}\,\sim\,15''$ in FWHM. Compared to the SCUBA observation of M82~\citep{leeuw-apj-2009} at 450$\,\mum$, the estimated distribution only roughly describes the double-peaked star--forming nucleus of M82. However, more detailed mathematical description of the intensity profile will introduce additional variables into the correction. Because the goal of this work is to correct the spectrum based on the limited information one can derive from the FTS observation, we retain the simple exponential profile. A comparison between our semi-extended corrected spectrum~(red and blue solid lines) and the point-source calibrated spectrum~(red and blue dashed lines) is shown in Figure~\ref{m82_spec}. Since the emission lines from $^{12}$CO at $J=4-3$ to $J=13-12$ are among the most important features in the FTS spectrum, the $^{12}$CO spectral line energy distribution~(SLED) is plotted in Fig.~\ref{m82_sled}. In this figure, our semi-extended corrected SLED is compared with the results from \citet{kamenezky-apj-2012}, where the spectrum was corrected by applying a source-beam coupling factor derived by convolving the M82 SPIRE photometer $250\,\mum$ map with appropriate profiles to produce the continuum light distribution seen with the FTS~\citep{panuzzo-aa-2010}. In Figure~\ref{m82_sled}, the CO line fluxes from the semi-extended corrected data were measured from the unapodized spectrum with a sinc function superposed on a parabola at a 20~$\mathrm{GHz}\,$ range centered at the redshifted frequency of each CO line. The measured line fluxes were multiplied by a factor of 0.55, which is the ratio of a simulated exponential profile with $r_{e}=11''$ after and before convolution by a Gaussian beam profile with FWHM$\,=42''$ at its central $9.5''\times9.5''$ region.
		}
		
		%  Section 5.1, Paragraph 4
		\par{
		As shown in Figure~\ref{m82_sled}, the semi-extended corrected line fluxes~(red) and the line fluxes measured and corrected by using the photometry map as a reference~(cyan) agree with each other within their error bars, even though the semi-extended corrected spectrum is based on a simplified circular exponential profile. It is interesting to note that the line fluxes for $J=8-7$, close to $959<\nu<989$~GHz, from the two spectra overlap in Figure~\ref{m82_sled}. For transitions higher than $J=8-7$, the semi-extended corrected line fluxes are higher than the line fluxes measured and corrected by using the photometry map as a reference, and for transitions between lower excitation levels than $J=8-7$, the semi-extended corrected line fluxes are slightly lower. This is due to the assumed circular exponential profile with $r_{\mathrm{e}}=11''$ has a different distribution than the photometry map.
		}
	
	% Section 5.2
	\subsection{Sgr B2}\label{sgrb2}

		% Section 5.2, Paragraph 1
		\indent\par{
		Sagittarius B2 (Sgr B2) is a giant molecular cloud at a distance of $\sim$8.3$\pm$0.4 Kpc \citep{Kerr86, Ghez08, Gillessen09, Reid09}, and located in the Central Molecular Zone at $\sim$120 pc from the Galactic Center \citep{Lis90}. It contains three main compact cores, Sgr B2(N), Sgr B2(M) and Sgr B2(S) distributed from north to south, associated with massive star formation. The FTS observed Sgr~B2 on 2011-02-27~(OD 655, ObsID 1342214843) in single pointing observations towards the (N) and (M) positions. Full details of these spectra are presented and analyzed by \citet{Etxaluze13}.
		}

		% Section 5.2, Paragraph 2
		\par
\begin{figure}
			\centering
			\includegraphics[height=7.5cm]{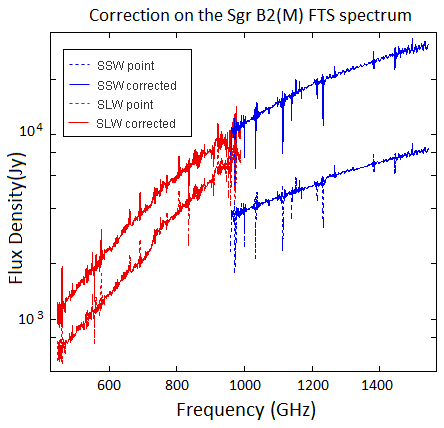}
			\caption{Correction of the Sgr B2(M) FTS spectrum. The dashed spectrum shows the point source calibration of the Sgr B2(M) spectrum. The solid line represents the Sgr B2(M) spectrum corrected with the semi-extended correction.}
			\label{SgrB2M}
		\end{figure}

{
		Figure~\ref{SgrB2M} shows the unapodized spectra at the position of Sgr~B2(M) calibrated as a point source (dashed) and after the application of our semi-extended correction. Assuming a Gaussian emission profile, the best continuum matching for the spectral continuum levels in the overlap region between SSW and SLW is obtained with FWHM$\,= 30''$ (this is equivalent to a radius of $\sim$0.62 pc). For the purpose of comparing the integrated line fluxes at the same spatial resolution, the resulting spectrum is convolved to a Gaussian beam with FWHM$\sim43''$, equivalent to a radius of $\sim$0.8 pc.
		}

		%  Section 5.2, Paragraph 3
		\par{
		The Sgr B2(M) spectral continuum, once corrected, provides a good measurement of the dust spectral energy distribution (SED) and allows us to estimate the luminosity, the mass and the molecular hydrogen column density of the source. These results obtained with the semi-extended corrected spectrum can then be compared with the results of previous studies.
		}

		% Section 5.2, Paragraph 4
		\par{
		The total integrated continuum intensity inside the beam with FWHM$\,=43''$ provides a luminosity of $L_{\rm FIR}= (5\pm1)\times10^6$ L$_{\odot}$, which is in good agreement with that measured by \citet{Goldsmith92}: $6.3\times 10^6$ L$_{\odot}$. The far-infrared luminosity requires several young O-type stars as power sources, which are deeply embedded in the star forming cores \citep{Jones08}.
		}

		% Section 5.2, Paragraph 5
		\par{
		The dust mass and the molecular hydrogen column density of the core is calculated as \citep{Deharveng09,Hildebrand83}:
		
			% Equation 19
			\begin{equation}
				M_{\rm d}=\frac{F_{250\mu m} d^2}{\kappa_{250\mu m}{B}_{250\mu m}(T_{\rm d})}\hspace{0.5cm}(\mathrm{M_{\odot}})
				\label{eq:Mass}
			\end{equation}

		\noindent where $F_{250\mu m}$ is the total intensity at 250~$\mu$m, {\it d} is the distance ($\sim8.5$ kpc), ${B}_{250\mu m}(T_{\rm d})$ is the blackbody intensity at 250~$\mu$m for a dust temperature $T_{\rm d}\sim 36.5$~K \citep{Etxaluze13}, and $\kappa_{250\mu m}=5.17$ cm$^2$g$^{-1}$ is the dust opacity \citep{Li01}.
		}

		% Section 5.2, Paragraph 6
		\par{The H$_2$ column density is given as: 

			% Equation 20
			 \begin{equation}
				N({\rm H_2})=\frac{\chi_{\rm d} M_{\rm d}}{2.3 m_{\rm H} d^2 \Omega_{\rm beam}}\hspace{0.5cm}(\mathrm{cm^{-2}})
				\label{eq:sgr_nh2}
			\end{equation}

		\noindent where we assumed a gas-to-dust ratio $\chi_{\rm d}=100$, $m_{\rm H}$ is the mass of a hydrogen atom and $\Omega_{{\rm beam}}$ is the beam solid angle. The H$_2$ column density derived is $N(\rm H_2)\sim 5\times 10^{24}$ cm$^{-2}$. The total mass of the core is: $M_{\rm d}=2300$ M$_{\odot}$. This mass is lower than the total masses determined by \citet{Qin11}: $\sim$3300 M$_{\odot}$, and larger than the mass calculated by \citet{Gaume90}: $\sim$1050 M$_{\odot}$. The derived physical quantities indicate that the corrected spectrum is generally in agreement with results obtained in previous studies of Sgr B2(M). If using an uncorrected spectrum, despite the apparent discontinuity at the overlap bandwidth, which makes fitting a blackbody continuum to the spectrum difficult, the derived value of $M_{d}$ would be $\sim900\,M_{\odot}$. This value is smaller than the previously measured dust mass in Sgr B2(M), leading to a possible underestimation of the dust mass in Sgr B2(M).
		}

		% Figure 9: Sgr B2(M) spectrum

% Section 6
\section{Conclusion}\label{discussion}

	% Section 6, Paragraph 1
	\indent\par{
	In this paper, we have shown that care must be exercised in the interpretation of spectral images produced by the SPIRE FTS as the spectra depend on both the frequency dependent beam and on the intrinsic source structure. While the beam profile can be determined from spectral scans of a point source, the effect of the source structure is less easy to model. However, since the beam profile in the overlap frequency range between the two bands employed by the SPIRE FTS is significantly different, this range provides a useful diagnostic on the source intensity distribution. For example, an abrupt change in the measured spectral intensity from one band to the other can often be directly related to the angular extent of the source.
	}

	% Section 6, Paragraph 2
	\par{
	An empirical method to correct for the effects induced by the discontinuity in beam size and shape for single-pointing sparse-sampling observations (Section~\ref{sec_semi}, Equation~\ref{source_correction}) has been presented. Alternatively, if a sufficiently simple analytic description of the source structure exists, a crude measure of the source extent can be derived from the SPIRE FTS observation by minimizing the intensity difference between the two bands. This work has now been included as an interactive tool in version 11 of the \textsl{Herschel} Interactive Processing Environment \citep[HIPE; ][]{ott-hipe} that is used for \textsl{Herschel} data processing.
	}

	% Section 6, Paragraph 3
	\par{
	Two examples have been presented which illustrate both the power of the technique and its limitations. Fundamentally, the tailored matching of the intensity in the overlap between the two bands can provide information on the angular extent of the source emitting continuum photons at those frequencies ($959<\nu<989~\mathrm{GHz}$). However, caution should be exercised in extrapolating this source structure to the interpretation of the remaining spectrum. This extrapolation may be valid in some cases, but, in general, the physical distribution of individual components depends on local conditions, e.g., density, temperatures of dust and gas, and chemical balance.
	}

	% Section 6, Paragraph 4
	\par{
	One of the primary advantages of Fourier transform spectrometers is their ability to provide intermediate spectral resolution over a relatively broad spectral range. In general, however, the spectral range is too large to be covered by feedhorn coupled detectors operating in single mode, as is the case for most heterodyne spectrometers, e.g. \textsl{Herschel} HIFI~\citep{degraauw-aa-2010}. The result is that the beam profile for an FTS exhibits a complex dependency on frequency as additional modes, propagated by the waveguides, couple to the instrument and ultimately form the beam on the sky. Complete solutions to Maxwell's equations for incident radiation propagating through an instrument tend to be extremely challenging and although ``quasi-optical'' approaches have been developed to simulate complex instruments~\citep{osullivan-spie-2009}, in practice the frequency dependent beam profile must be determined from spectral observations of a point source~\citep{makiwa-prep-2013}.
	}

	% Section 6, Paragraph 5
	\par{
	While it is in principle possible to illuminate a detector array using reflective camera optics, which provides a well-defined beam profile (i.e., without the use of feedhorns; as is done for example with FTS-2~\citep{naylor-spie-2003}, the Fourier spectrometer developed for use with the SCUBA-2 camera~\citep{holland-spie-2006}), it is difficult to control stray light in such direct imaging applications. While stray light is less of a concern for ground based instruments, whose sensitivity is limited by the photon flux from the warm telescope and atmosphere, control of stray light is of critical importance in cryogenic far-infrared space astronomy missions currently being proposed~\citep{swinyard-expa-2009}. To exploit the sensitivity of state-of-the-art detectors on these missions stray light must be controlled using feedhorn coupled detectors. The SAFARI instrument~\citep{roelfsema-spie-2012} under development for the SPICA mission is an imaging FTS employing feedhorns and will encounter a similar frequency dependent beam profile and require a similar analysis for semi-extended sources as that described in this paper.
	}

\begin{acknowledgement}

\par{SPIRE has been developed by a consortium of institutes led by Cardiff University (UK) and including Univ. Lethbridge (Canada); NAOC (China); CEA, LAM (France); IFSI, Univ. Padua (Italy); IAC (Spain); Stockholm Observatory (Sweden); Imperial College London, RAL, UCL-MSSL, UKATC, Univ. Sussex (UK); and Caltech, JPL, NHSC, Univ. Colorado (USA). This development has been supported by national funding agencies: CSA (Canada); NAOC (China); CEA, CNES, CNRS (France); ASI (Italy); MCINN (Spain); SNSB (Sweden); STFC (UK); and NASA (USA).}
\\\\
\par{RW would like to thank Dr. J. Bock, F. Galliano, S. Hony, J. Kamenetzky, and, C. Wilson for their helpful discussions and comments on this work. ME thanks ASTROMADRID for funding support through the grant S2009ESP-1496, the Spanish MINECO (grant AYA2009-07304) and the consolider programme ASTROMOL: CSD2009-00038. GM, DN and MvdW acknowledge support from NSERC.}
\end{acknowledgement}

%\appendix
%\input{sect_note}

\end{document}